\documentclass[epsfig,useAMS,usenatbib]{mn2e}
\usepackage{epsfig,graphics,latexsym,subfigure}
\newcommand{\mc}{\multicolumn}

\newcommand{\gtsim}{\mbox{{\raisebox{-0.4ex}{$\stackrel{>}{{\scriptstyle\sim}}
$}}}}

\newcommand{\vvmax}{$V/V_{\rm max}\,$}

\begin{document}

\title[The cosmic evolution of low-luminosity radio sources from the SDSS DR1]{The cosmic evolution of low-luminosity
radio sources from the SDSS DR1}

\author[Clewley \& Jarvis]{L. Clewley\thanks{clewley@astro.ox.ac.uk} \& Matt J.\,Jarvis\thanks{mjj@astro.ox.ac.uk}\\
Astrophysics, Department of Physics, Keble Road, Oxford, OX1 3RH, UK }
\maketitle
 
\begin{abstract}
In this paper we constrain the evolution in the comoving space
density for low-luminosity (typically FRI) radio sources as a function
of cosmic epoch by matching deep radio surveys (WENSS, FIRST and NVSS)
with the Sloan Digital Sky Survey (SDSS). This results in 1014 matched
radio sources covering an effective area of 0.217~sr, 686 with
$L_{325 \rm MHz} < 10^{25}$~W~Hz$^{-1}$~sr$^{-1}$, which is an order of
magnitude larger than any previous study at these luminosities. Using the non-parametric
\vvmax test we show that low-luminosity radio sources evolve
differently to their more powerful, predominantly FRII,
counterparts. Indeed, we find that the lower luminosity population is
consistent with a constant comoving space density with redshift, as
opposed to the strong positive evolution for the more powerful
sources.
\end{abstract}

\section{Introduction}\label{sec:intro}
The evolution of radio-loud active galactic nuclei (AGN) has been a
difficult quantity to pin down for all but the most powerful sources
up to moderate redshifts. These sources appear to trace the most
massive galaxies at all cosmic epochs (e.g. Jarvis et al., 2001;
Willott et al., 2003) and so understanding their evolution is of
strategic importance to astronomy. Further, radio sources make
particularly useful probes of galaxy and AGN evolution because radio
emission is not affected by dust obscuration.  The evolution in the
space density of the radio source population is therefore a crucial
element in our understanding of the AGN phenomena.

It has been suggested that the radio AGN population is composed of
different types of objects. The more powerful sources show a high
surface brightness at the ends of a double lobed structure
(Fanaroff-Riley type II; hereafter FRII; Fanaroff \& Riley 1974)
whereas the less powerful sources have high surface brightness nearest
their centres (FRI). It is now well established that the FRII radio
sources evolve strongly between $z \sim 2 \rightarrow 0$ (Longair
1966; Dunlop \& Peacock 1990) with a decrease in the comoving space
density of $\sim 1000$ since $z \sim 2$.  However, very little is
known about the lower-luminosity FRI radio sources. Various groups
have tried to understand how the lower-luminosity radio sources evolve
with redshift. The conclusions drawn from these studies vary. For
instance, Jackson \& Wall (1999) find evidence for possible negative
evolution. Laing, Riley \& Longair (1983) find, from 17 sources in the
complete 3CRR sample, evidence for no-evolution. Similarly, a study by
Waddington et al. (2001) of 72 sources also suggests little evidence of
evolution in the space density between z = 0 and z = 1 in the
low-luminosity regime. However, Snellen \& Best (2001) find, from two
objects in the Hubble Deep Field and the flanking fields, evidence for
positive evolution. This fundamental discrepancy arises because of
the observational difficulty in establishing a large complete
flux-density-limited radio sample at faint fluxes due to the vast
overall increase in radio source counts.

The problem of small number statistics can now be overcome with large
multiwavelength surveys. The advent of the large radio surveys such as
the 325 MHz Westerbork Northern Sky Survey (WENSS); the Faint Images
of the Radio Sky at Twenty-cm (FIRST) and the 1.4 GHz NRAO VLA Sky
Survey (NVSS) along with large scale optical surveys such as the Sloan
Digital Sky Survey Data Release 1 (SDSS DR1) allow us to probe the
low-luminosity radio source population in a statistically complete
manner. These surveys, when combined, provide the most efficient
method of investigating the evolution of the faint radio source
population as a function of cosmic epoch. The limiting factors are the
flux-density limits of the radio surveys and the depth of the optical
data. The WENSS and FIRST surveys can probe
a radio source with $L_{325 \rm MHz} = 10^{25}$ W~Hz$^{-1}$~sr$^{-1}$
(roughly at the FRI/FRII break luminosity) at a flux limit of $S_{325} =
40$~mJy up to $z \sim 1$. The most limiting constraint comes from an
upper limit in the optical flux from the SDSS (we adopt $i < 21$).

In this paper we investigate the evolution in the comoving space
density of low-luminosity radio sources with cross-matched catalogues
formed from the WENSS, FIRST and NVSS radio surveys and the recently
released photometric survey SDSS DR1. In \S2 we describe the selection
of the radio sources and their matching to the SDSS DR1. \S3 contains
details of the calculation of the photometric redshifts. In \S4 the
evolution in the comoving space density of radio sources is tested via
the \vvmax statistic. In \S5 we present the results and \S6 is a
summary of our conclusions.

We define radio spectral index $\alpha_{\rm rad}$ as 
$S_{\nu} \propto \nu^{-\alpha_{\rm rad}}$, where $S_{\nu}$ is the
flux-density at frequency $\nu$. We assume throughout that $H_{\circ}=70~
{\rm km~s^{-1}Mpc^{-1}}$ and a $\Omega_ {\rm M}=0.3$, $\Omega_
{\Lambda}=0.7$ cosmology.

\section{Sample Selection}
\begin{figure}
\centering{ \scalebox{0.35}[0.2333]{
\includegraphics*[20,140][600,700]{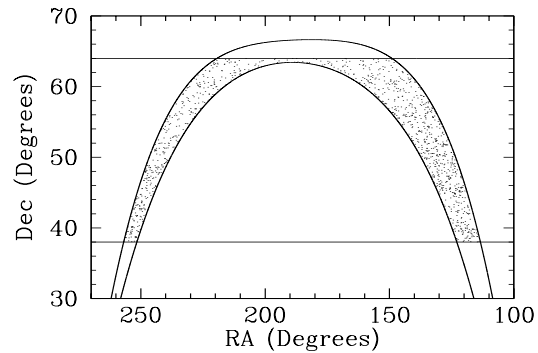} }}
\caption{The survey area of the matched radio and photometric sample
  covering a total area of survey is 713 square degrees (0.217 sr). The
  curves are the area covered by the SDSS DR1 which are bounded by
  FIRST at a declination of 64$^{\circ}$ and WENSS at 38$^{\circ}$. There are
  1014 matched $\alpha_{\rm rad} > 0.5$ radio sources in the survey (shown as
  dots).}
\label{survey_area}
\end{figure}

We select a sample area that is sufficiently large to increase the
number of low-luminosity, predominantly FRI, radio sources from $\sim$
tens to about 1000. We choose the continuous area of the northern SDSS
DR1 sample which is bounded by FIRST at a declination of 64$^{\circ}$ and
WENSS at 38$^{\circ}$. The shaded are in Fig.~\ref{survey_area} shows the resulting area
of our survey, covering 713 square degrees (0.217~sr).

\subsection{Radio source selection}
We wish to select radio galaxies where the optical light is dominated
by the host galaxy, rather than quasars, and the radio emission may
undergo Doppler beaming effects along the line-of-sight (e.g. Jarvis
\& Rawlings 2000). Therefore, we start with the low-frequency WENSS
radio survey that will preferentially select optically thin steep
spectrum lobe emission rather than optically thick flat spectrum
(possibly beamed) emission.

The WENSS radio sources are matched to NVSS and FIRST radio surveys,
bounded by the SDSS DR1 photometric survey. We summarise each of these
surveys only briefly here - details can be found at each of their
respective websites and references therein.

The WENSS survey (Rengelink et al. 1997;
http://www.strw.leidenuniv.nl/wenss/) covers 3.14 sr north of
+30$^{\circ}$ declination at 325 MHz with resolution of 54'' and a
flux-density completeness limit of approximately 18 mJy
(5$\sigma$). We choose a conservative flux-density limit of 40~mJy
which provides the limit for this survey. The NVSS radio survey
(Condon et al. 1998; http://www.cv.nrao.edu/nvss/) is a 1.4 GHz
continuum survey covering 10.3 sr of sky north of -40$^{\circ}$
declination. The completeness limit is about 2.5 mJy and the images
have a 45'' resolution and a nearly uniform sensitivity. The FIRST
survey (Becker, White \& Helfand 1995; White et al. 1997;
http://sundog.stsci.edu) is at 1.4 GHz and covers 2.75 sr at
-10$^{\circ} <$ declination $< 64^{\circ}$ with resolution of 5'' and
a 1~mJy source detection threshold.

\subsection{Radio source matching}
We cross-match the WENSS and NVSS catalogue positions taking all
matches with positional offsets of $\leq$ 40'' as candidate radio
detections. (We use the low-resolution NVSS survey over FIRST to
determine 1.4~GHz flux-densities as some of the flux may be resolved
out by the higher-resolution FIRST survey.) Increasing the distance to
$\leq$ 45'' increases the number of matched targets by only $\sim$
1\%. At this distance the probability of choosing a NVSS source within
an arbitrary position is 2\% whereas for the WENSS catalogue this is
0.8\%, so we expect the contamination to be at the level of $\sim 1\%$
and the 40'' offset is considered optimal.

We calculate the radio spectral index, $\alpha_{\rm rad}$, for each
source using the flux-density at frequencies 325 MHz and 1.4 GHz from
the matched radio sources in the low-resolution WENSS and NVSS surveys
respectively. In order to select against quasars and other non-lobe
dominated sources we choose $\alpha_{\rm rad}$ $>$ 0.5.

The NVSS coordinates of this cross-matched sample are further matched
to the high resolution radio galaxy catalogue, FIRST.  Again we choose
a positional offset of $\leq$ 40''. We do this for two reasons: (i)
the increased resolution of FIRST allow us to improve the positional
accuracy of the matched WENSS-NVSS catalogue, thus reducing the number
of mis-identified sources with the photometric survey and (ii) the
FIRST survey image cutouts allow an accurate visual check on the
matching procedure to the SDSS DR1. The matched WENSS-NVSS-FIRST
yield a total of 7898 matched radio targets in the sample area to be
then matched to the photometric survey.

\subsection{Photometric matching}
We cross-match the SDSS DR1 catalogue with the FIRST matched
positions, with positional offsets of $\leq$ 5''. We use the SDSS
modified Petrosian magnitudes (Petrosian, 1976; Strauss et al.,
2002). In this magnitude system the aperture size depends on the shape of the galaxy's radial
surface brightness profile but not its amplitude. The disadvantage of
this system is that the photometric measurements become
increasingly noisy for $r$ fainter than around 20. The other option is
to use SDSS model magnitudes which are less noisy for the faint
galaxies but have serious systematics for galaxies $r$ $<$ 20. The
SDSS DR1 quote point spread function magnitude limits (95\% detection repeatability for
point sources) of 22.0, 22.2, 22.2, 21.3, 20.5 in $u, g, r, i, z$
respectively with the errors in the magnitudes $\leq$ 3\%. For Petrosian
magnitudes the reported errors are considerably larger than this. To
ensure our sample is complete, and to minimise the photometric errors,
we take a colour cut of $i <$ 21. This effectively limits our analysis
to z $\leq$ 1, assuming the host galaxies are typically $\sim 3
L^{\star}$ (McLure et al. 2004) - which makes it the most limiting
factor in the survey. Additionally, we use the SDSS morphological
classification and select only non-stellar sources thus selecting
against quasars.

\begin{figure}
\centering{ \scalebox{0.35}[0.2333]{
\includegraphics*[20,140][600,700]{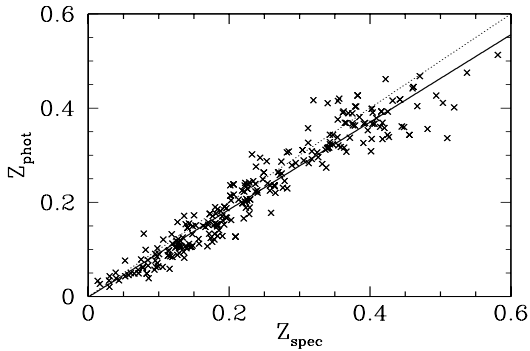} }}
\caption{Photometric redshifts versus spectroscopic redshifts for 265
 radio sources in our sample with spectroscopic redshifts. The ANN$z$
 code is trained on $\sim 120,000$ galaxies from the SDSS. A linear
 least-squares fit to the sample with z $\leq$ 0.5 gives $Z_{\rm phot}
 = Z_{\rm spec}*0.943 -0.003$, shown as a solid line. The dotted line
 is $Z_{\rm phot} =  Z_{\rm spec}$. }
\label{fig_annz}
\end{figure}

We overplot the optical matches from SDSS DR1 onto the radio images
from FIRST. In most cases the radio source is unambiguous. However,
careful visual inspection of the images reveals that 8\% of the
sources with optical identifications within 1' are not easily
associated with a core of a radio galaxy. The majority of these cases
are due to multiple identifications along the radio axis, which
presumably arises from the well known alignment effect of the optical
and radio emission (e.g. Eales et. al. 1997; Best, R\"ottgering \&
Longair, 2000; Inskip et al. 2002). As the narrow-line luminosity is
correlated with radio luminosity (e.g. Willott et al. 1999) then this
effect will be most prominent in the more powerful FRII subset. Thus
we are confident that the survey as a whole is $>$ 90\% complete, and
the low-luminosity sources - which are our main concern in this paper
- have a higher completeness. These selections result in 1014 radio
sources with associated photometric data.

We incrementally alter the search radius of the optically matched
FIRST and SDSS DR1 radio sources from 5 arcsec to 2 arcsec in order to
investigate the effect on the contamination of mis-identified sources
and the completeness of the sample. Ivezi{\' c} et al. (2002) discuss
this problem in detail. These authors find that with a 1 arcsec search
radius the sample is 72\% complete and 1.5\% contaminated. Increasing
the search radius to 3 arcsec results in $>$ 99\% completeness and 9\%
contamination. We expect the contamination in our survey to be
significantly less than that found by Ivezi{\' c} et al. as
the matching of each radio source is visually inspected by
overplotting the SDSS DR1 photometric data on maps of the FIRST radio
sources, whereas the Ivezi{\' c} et al. study was automated.

\section{Photometric redshifts}\label{sec:photz}
Redshift information is crucial in determining the evolution in the
comoving space density of radio sources. Gaining spectroscopic
redshift information for all of the radio sources in our sample would
be extremely time consuming on any telescope. However, the five-band
photometry from the SDSS DR1 allows us to use photometric redshifts to
make reasonably accurate estimates of the whole sample. We use two
publicly available codes to do this. The first is {\sc Hyper}{\it z}
(Bolzonella, Miralles \& Pell, 2000), which uses galaxy templates and
various degrees of reddening to determine the most likely redshift. We
assume {\it a priori} that the host galaxies of our radio sources are
either elliptical galaxies or bulge dominated spirals (e.g. McLure et
al. 2004), and as such only use the corresponding template spectra.

The second code (ANN$z$; Collister \& Lahav 2004) estimates
photometric redshifts by employing neural networks which require a
large training set of galaxies for which the redshifts are already
known. Collister \& Lahav demonstrate the effectiveness of their
method using SDSS DR1 and compare it to the {\sc Hyper}{\it z}
technique. They find that when a large training set is available the
ANN$z$ technique is more robust than {\sc Hyper}{\it z} method with
smaller random and systematic errors.

We use the large sample of galaxies with spectroscopic redshifts and
five-band photometry from the SDSS DR1, which comprises $\sim 120,000$
galaxies. In Fig.~\ref{fig_annz} we show the accuracy of the ANN$z$
technique for targets in our sample with spectroscopic
measurements. The random error on the photometric redshifts is
typically $\Delta z \sim 0.03$ at $z < 0.5$. Inspection of
Fig.~\ref{fig_annz} reveals a slight systematic error. We discuss the
effect of this systematic underestimate of the photometric redshifts
in \S\ref{sec:vvmax}.

Beyond $z~\gtsim~0.5$ the errors in the ANN$z$ analysis become larger,
predominantly due to the fact that the majority of the galaxies
observed in the spectroscopic training sample have $i < 19.5$. The
relative dearth of high-redshift sources in the spectroscopic training
set fainter than this limit means that ANNz is not the ideal
photometric redshift code for this region of parameter
space. Therefore we use {\sc Hyper}{\it z} estimates for sources with
$i > 19.5$.

\section{The V/V$_{\rm max}$ statistic}\label{sec:vvmax}
In order to assess the evolution in the comoving space density of
radio sources we use the non-parametric $V/V_{\rm max}$ method
(Schmidt 1968; Rowan-Robinson 1968). This technique is useful as we do
not have to make any {\it a priori} assumptions as to the form of any
evolution. It also allows us to easily incorporate additional
selection criteria. Indeed this is critical for this study as we have
an optical magnitude limit of $i < 21.0$~mag.

Under the null hypothesis of a uniform distribution, the value of
\vvmax will be uniformly distributed between 0 and 1. For such a
sample the mean value is $\langle$\vvmax$\rangle = 0.5 \pm
(12N)^{-\frac{1}{2}}$, where $N$ is the number of objects in the
sample. A value of $\langle$\vvmax $\rangle > 0.5$ indicates that the
sources are biased towards larger distances, or an increasing space
density with redshift; whereas $\langle$\vvmax $\rangle < 0.5$
indicates a deficit of high-redshift sources, or a decline in the
space density evolution. The total fraction of sources for which Vmax
is set by the optical selection criteria is 35\%. This is split into
13\% of the low-luminosity sources and 83\% of the high-luminosity
sources.
\begin{table}
\begin{center}
{\caption{\label{tab:samples} 
\vvmax\, for 8 bins in 325~MHz radio luminosity.}}
\begin{tabular}{c|c|c|c|c|c}
\hline\hline 
\mc{1}{c|}{$L_{325 \rm MHz }$} & \mc{1}{c|}{$N$} & \mc{1}{c|}{\vvmax} &
\mc{1}{c|}{$1\sigma$} & \mc{1}{c|}{$<z>$} & \mc{1}{c|}{\vvmax($z_{\rm adj}$)} \\
\hline\hline
23-23.5  &  55  &   0.43 &   0.04 &   0.11 & 0.42 \\
23.5-24  & 125  &   0.42 &   0.03 &   0.18 & 0.43 \\
24-24.5  & 241  &   0.52 &   0.02 &   0.31 & 0.52 \\
24.5-25  & 265  &   0.45 &   0.02 &   0.45 & 0.46 \\
25-25.5  & 192  &   0.54 &   0.02 &   0.66 & 0.53 \\
25.5-26  &  86  &   0.62 &   0.03 &   0.79 & 0.63 \\
26-26.5  &  31  &   0.70 &   0.05 &   0.96 & 0.72 \\
26.5-27  &  13  &   0.62 &   0.08 &   0.91 & 0.64 \\
\hline\hline
\end{tabular}
\end{center}
\end{table}
To incorporate the additional optical selection into the \vvmax
statistic we need to know the redshift at which the galaxy fades to $i
> 21$~mag. This is found by fitting the multi-colour data with various
galaxy templates generated from the GISSEL96 synthetic galaxy spectra
library of Bruzual \& Charlot (2003). We use ten different templates
of an evolving elliptical galaxy, with the redshift given by the
photometric redshift techniques described in
\S~\ref{sec:photz}. The spectral fit to the data is minimised
with a simple $\chi^{2}$ fitting routine and subsequently inspected by
eye (this also acted as a further consistency check of the photometric
redshifts). We find 95\% of the sources are well fitted by at least
one of the galaxy templates. The templates are then used to determine
the redshift at which the source would drop out of the sample as a
consequence of both cosmological dimming and $k$-correction, and this
is used as the $V_{\rm max}$ if it is found to be lower than the
$V_{\rm max}$ determined from the radio data alone.

We conduct the \vvmax test with and without those sources with poor
fits to the multi-colour data. The inclusion of the sources with
poorer fits leads to a decrease in the \vvmax estimate by $\sim 0.01$
at $L_{325 \rm MHz} < 10^{25}$~W~Hz$^{-1}$~sr$^{-1}$, and $\sim 0.02$
at higher luminosities, thus we do not deem it to be an important
uncertainty in our analysis. As well as the statistical errors, due to
the size of the sample which decrease as $(12N)^{-\frac{1}{2}}$, we
investigate three other sources of error. First, the redshift estimate
for each radio source has an associated uncertainty which is
principally dependant on the magnitude errors and hence is a function
of redshift. We use Monte-Carlo methods to assess the size of this
source of error. We derive 1000 samples drawn from the true sample
based on Gaussian deviations of the reported redshift error. Second,
we re-derive the \vvmax statistic for the sample after correcting for
the systematic photometric underestimate seen in
Fig.~\ref{fig_annz}. Finally - as we discuss in \S2 - we investigate
the effect of contamination of mis-identified sources on the \vvmax
statistic by altering the search radius of FIRST-SDSS DR1 matched
sources. We re-derive the \vvmax statistic for search radii of 2, 3 and
4 arcsec separations.
\section{Results}
\begin{figure}
\centering{ \scalebox{0.48}{
\includegraphics*[72,56][594,452]{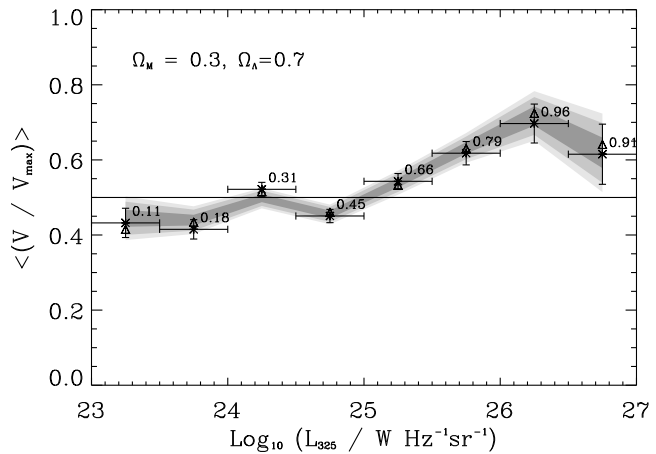} }}
\caption{The \vvmax statistic for low-luminosity ($L_{325 \rm MHz} <
  10^{27}$~W~Hz$^{-1}$~sr$^{-1}$) radio sources. Crosses denote the
  mean \vvmax statistic at a given redshift provided in
  Table~\ref{tab:samples}.  The shaded regions are the 68\%, 95\% and
  99\% (dark to light shading respectively) confidence intervals based
  on Gaussian deviations about the reported redshift. The open
  triangles are the \vvmax statistic derived after correcting for the
  systematic underestimate in the photometric redshifts.}
\label{fig_rlf}
\end{figure}

In Fig.~\ref{fig_rlf} and Table~\ref{tab:samples} we show the mean
\vvmax statistic in 8 bins of radio luminosity in the range $10^{23} <
L_{325 \rm MHz} < 10^{27}$~W~Hz$^{-1}$~sr$^{-1}$. For each bin we
provide the mean redshift $\langle z \rangle$. The 1$\sigma$ error
bars are the statistical errors derived from the sample size. As we
discussed above the \vvmax statistic is affected by the errors in the
estimates of the redshift. Overplotted in Fig.~\ref{fig_rlf} are the
results of the Monte Carlo simulations at 68\%, 95\% and 99\%
confidence (dark to light shading respectively) based on Gaussian
deviations about the reported redshift. We also re-derive the \vvmax
statistic for the sample after correcting for the systematic
underestimate seen in Fig.~\ref{fig_annz}. These adjusted estimates
are provided in the last column of Table~\ref{tab:samples} and are
shown as open triangles in Fig.~\ref{fig_rlf}.

The 5\% of objects that cannot be satisfactorily fit with the galaxy
templates change the \vvmax estimate by 0.02 at most. An inspection of
the confidence limits in Fig.~\ref{fig_rlf} suggests that the errors
that arise from the uncertainty in the redshift are less significant
than the statistical errors. Further, the systematic photometric error
leads to a change in $<$\vvmax$>$ of $\sim 0.01$. We note the latter
two sources of error have the combined effect of moving the \vvmax
statistic towards \vvmax = 0.5 for the $L_{325 \rm MHz} < 10^{25}$~W~Hz$^{-1}$~sr$^{-1}$ sample and away from it for the higher luminosity sample.

\begin{figure}
\centering{\scalebox{0.48}{
\includegraphics*[72,56][594,452]{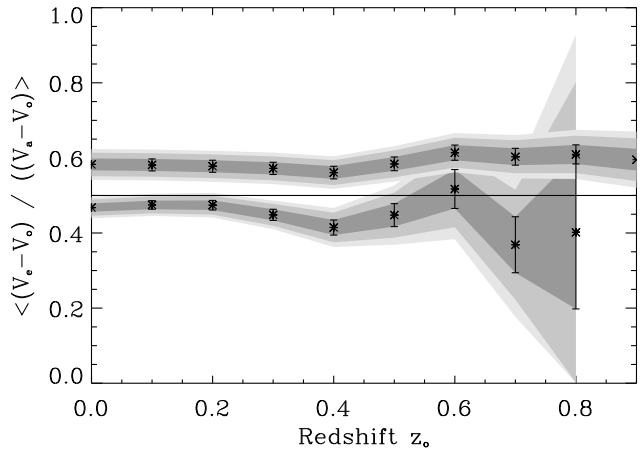}}}
\caption{The banded \vvmax statistic as a function of redshift for high-luminosity sources
  ($L_{325 \rm MHz} > 10^{25}$~W~Hz$^{-1}$~sr$^{-1}$; upper half) and low-luminosity radio
  sources ($L_{325 \rm MHz} < 10^{25}$~W~Hz$^{-1}$~sr$^{-1}$;
  lower half). In the plot V$_{e}$ is volume enclosed at the redshift of the source, V$_{a}$ is the volume which corresponds to the maximum redshift which the source can be seen and V$_{0}$ is the volume which corresponds to z$_{0}$. The shaded regions have the same meanings as Fig.~\ref{fig_rlf}.}
\label{fig_vvmax_z}
\end{figure}

The only evidence ($>2\sigma$) for a strong positive evolution comes
from the powerful radio sources with $L_{325 \rm MHz} >
10^{25}$~W~Hz$^{-1}$~sr$^{-1}$. It has long since been established
that these powerful, predominantly FRII sources, exhibit positive
evolution (i.e. \vvmax $>$ 0.5). However, the lower-luminosity
population does not follow this trend. In contrast to the strong
evolution indicated by the FRII sources, the lower luminosity sample
shows little evidence for evolution up to $z \sim 0.6$. This dichotomy
is illustrated further in Fig.~\ref{fig_vvmax_z} and Table
\ref{tab:vvmax_z}, where we show the banded \vvmax\, statistic, which
enables us to see the evolution in the \vvmax\, statistic as a
function of redshift (see Avni \& Bahcall 1980) for two bins in radio
luminosity [$L_{325 \rm MHz} > 10^{25}$~W~Hz$^{-1}$~sr$^{-1}$ (upper)
and $L_{325 \rm MHz} < 10^{25}$~W~Hz$^{-1}$~sr$^{-1}$ (lower)]. This
result remains unchanged if we alter the search radius of the
optically matched FIRST and SDSS DR1 radio sources. 

Table \ref{tab:vvmax_arcsec} shows the same banded \vvmax\, statistic in the
$L_{325 \rm MHz} < 10^{25}$~W~Hz$^{-1}$~sr$^{-1}$ bin as in Table
\ref{tab:vvmax_z} but for search radii of 4, 3 and 2 arcsec. The
banded \vvmax\, statistic remains consistent in each of the
separations but the relative levels of completeness fall to 80\%,
68\% and 57\% for 4, 3 and 2 arcsec separations respectively. However, due to
the consistency in \vvmax\, the completeness seems to be uniform in
all $z$ bins.

The evidence for the lack evolution in the low-luminosity sources is
broadly consistent with a radio survey by Waddington et al. (2001) which
uses 72 sources, 42 of which have accurate spectroscopic
redshifts. Examination of their Fig. 10 suggests no significant
evolution of sources at $0 < z < 1$ which is consistent with our lowest
luminosity bin of $P_{325 \rm MHz} \sim 10^{23}$ W~Hz$^{-1}$~sr$^{-1}$
shown in Fig.~\ref{fig_rlf}.

Jackson \& Wall (1999) found that for 26 sources at the lower
luminosity, $<$\vvmax$>$ = 0.314$\pm$0.057. If we calculate the mean
\vvmax\, statistic for our low-luminosity and high-luminosity sources
over all redshifts we find that the sample with $10^{23} < L_{325 \rm
MHz} < 10^{25}$~W~Hz$^{-1}$~sr$^{-1}$ yields $<$\vvmax$>$ =
0.47$\pm$0.01 for 686 sources. The sample $L_{325 \rm MHz} >
10^{25}$~W~Hz$^{-1}$~sr$^{-1}$ has a $<$\vvmax$>$ = 0.58 $\pm$
0.02. This is obviously calculated over a large range in redshift
which may not be informative regarding the evolution. Indeed two
samples with different $<$\vvmax$>$ do not necessarily have different
evolutions (Anvi \& Bahcall 1980). However it does indicate that the
low-luminosity sources are consistent with little or no evolution and
that the Jackson \& Wall (1999) values are ruled out at $> 5\sigma$. A
direct comparison between such studies and the one presented here is
made difficult because of the small sample size so that the samples
may be drawn from completely different distributions.

We also note that our results are consistent with Snellen \& Best
(2001) as their two sources are at $L_{325 \rm MHz} \sim
10^{25}$~W~Hz$^{-1}$~sr$^{-1}$ and as such are at the cross-over point
between the lower and higher luminosity sources in our analysis.
\begin{table}
\begin{center}
{\caption{\label{tab:vvmax_z} 
The banded \vvmax\, statistic as a function of redshift split at
$L_{325 \rm MHz} = 10^{25}$~W~Hz$^{-1}$~sr$^{-1}$.}}\begin{tabular}{c|c|c|c|c}
\hline\hline 
\mc{1}{c|}{$z$} & \mc{1}{c|}{$\langle \frac{V_{e} - V_{\circ}}{V_{a} -
    V_{\circ}} \rangle$} & \mc{1}{c|}{$1\sigma$} &
\mc{1}{c|}{$\langle \frac{V_{e} - V_{\circ}}{V_{a} -
    V_{\circ}} \rangle$} & \mc{1}{c|}{$1\sigma$} \\
& \mc{2}{c|}{($L_{325 \rm MHz} < 10^{25}$)}  & \mc{2}{c|}{($L_{325 \rm MHz} > 10^{25}$)}\\
\hline\hline
0.0 & 0.47 &   0.01  & 0.58 &   0.02    \\ 
0.1 & 0.47 &   0.01  & 0.58 &   0.02    \\
0.2 & 0.47 &   0.01  & 0.58 &   0.02    \\ 
0.3 & 0.45 &   0.01  & 0.57 &   0.02    \\
0.4 & 0.41 &   0.02  & 0.56 &   0.02    \\ 
0.5 & 0.45 &   0.03  & 0.58 &   0.02    \\ 
0.6 & 0.52 &   0.05  & 0.61 &   0.02    \\ 
0.7 & 0.37 &   0.07  & 0.60 &   0.02    \\ 
0.8 & 0.40 &   0.20  & 0.61 &   0.03    \\ 
0.9 &  -   &    -    & 0.60 &   0.03    \\
\hline\hline
\end{tabular}
\end{center}
\end{table}
\section{Summary and discussion}
We have conducted the first large scale analysis of the evolution in
the comoving space density of radio sources at the faint end of the
radio luminosity function. We match steep spectrum sources from the
WENSS, NVSS and FIRST radio catalogues to the SDSS DR1 photometric
catalogue which results in 1014 sources with redshift information
accurate to $\Delta z \sim 0.03$ at $z < 0.5$. Of these sources 686
sources have $L_{325 \rm MHz} < 10^{25}$~W~Hz$^{-1}$~sr$^{-1}$ and 328
have $L_{325 \rm MHz} > 10^{25}$~W~Hz$^{-1}$~sr$^{-1}$. For the
powerful sources, assumed to be predominantly FRII radio sources, we
are in good agreement with previous work based on spectroscopically
complete radio samples (e.g. Dunlop \& Peacock 1990; Willott et
al. 2001). However, for the lower luminosity population, we find that
the comoving space density remains approximately constant with
increasing redshift, and thus these objects do not follow the trend of
their more powerful counterparts.

This is in rough agreement with what one would expect by combining the
information from both the radio luminosity function derived from a 
brighter, complete flux-density limited sample (e.g. Willott et
al. 2003) and the low-frequency source counts. We rule out a strong
negative evolution, as suggested by Jackson \& Wall (1999) at $>
5\sigma$. We also find no evidence for a significant increase in the
comoving space density of the sources with $L_{325 \rm MHz} <
10^{25}$~W~Hz$^{-1}$~sr$^{-1}$. However, it
is difficult to assess the significance of any possible dichotomy in
the evolution of the low-luminosity and the high-luminosity sources
with our sample alone, as we probe completely different ranges in
redshift, due to redshift -- luminosity degeneracy inherent to
flux-density limited samples. In comparison to the combined 3CRR, 6CE
and 7CRS samples of Willott et al. (2001) which covers a combined area
of $\sim 4.3$~sr with virtually complete redshift information, and has
many powerful radio sources at $z < 0.5$, our analysis shows that the
lower-luminosity population is unlikely to have evolved in such a
dramatic way. Whether this is due to the evolution in comoving space
density correlating with radio luminosity or with FRI/FRII morphology
is impossible to determine with the resolution of the FIRST survey
used in this analysis. Access to further high-resolution radio data
would fully resolve this issue.
\begin{table}
\begin{center}
{\caption{\label{tab:vvmax_arcsec} 
The banded \vvmax\, statistic for the low luminosity sources ($L_{325
  \rm MHz} < 10^{25}$~W~Hz$^{-1}$~sr$^{-1}$) as derived in Table 2 but
for FIRST-SDSS DR1 matching 
radii of 4, 3 and 2 arcsec.}}\begin{tabular}{c|c|c|c|c|c|c}
\hline\hline 
\mc{1}{c|}{$z$} &
\mc{1}{c|}{4''} &
\mc{1}{c|}{1$\sigma_{4''}$} & \mc{1}{c|}{3''} &
\mc{1}{c|}{1$\sigma_{3''}$}  & \mc{1}{c|}{2''} &
\mc{1}{c|}{1$\sigma_{2''}$} \\
\hline\hline
   0.0  &  0.47 & 0.01  &  0.47  &  0.01  &  0.48 & 0.01 \\
   0.1  &  0.48 & 0.01  &  0.48  &  0.01  &  0.49 & 0.02 \\
   0.2  &  0.48 & 0.01  &  0.48  &  0.02  &  0.48 & 0.02 \\
   0.3  &  0.45 & 0.02  &  0.45  &  0.02  &  0.45 & 0.02 \\
   0.4  &  0.42 & 0.02  &  0.41  &  0.02  &  0.40 & 0.03 \\
   0.5  &  0.45 & 0.03  &  0.41  &  0.04  &  0.42 & 0.04 \\
   0.6  &  0.52 & 0.06  &  0.45  &  0.06  &  0.42 & 0.07 \\
   0.7  &  0.35 & 0.09  &  0.25  &  0.10  &  0.23 & 0.13 \\
   0.8  &  0.74 & 0.29  &   -    &   -    &   -   &  -   \\
\hline\hline
\end{tabular}
\end{center}
\end{table}

\section*{Acknowledgements}
LC and MJJ are funded by PPARC PDRAs. The authors would like to thank
the referee, Jim Dunlop, for a careful reading of the manuscript and
helpful comments. We thank Steve Rawlings and Garret Cotter for useful
discussions and Adrian Collister for providing the source code of
ANN$z$.  This publication makes use of the material provided in the
WENSS, FIRST, NVSS and SDSS DR1 surveys. The WENSS project is a
collaboration between the Netherlands Foundation for Research in
Astronomy and the Leiden Observatory. NVSS and FIRST are funded by the
National Astronomy Observatory (NRAO) and is a research facility of
the U.S. National Science foundation and use the NRAO Very Large
Array. Funding for the creation and distribution of the SDSS Archive
has been provided by the Alfred P. Sloan Foundation, the Participating
Institutions, the National Aeronautics and Space Administration, the
National Science Foundation, the U.S. Department of Energy, the
Japanese Monbukagakusho, and the Max Planck Society. Further details
of the SDSS survey can be found on http://www.sdss.org/.


\begin{thebibliography}{99}
\bibitem{15}  Avni Y., Bahcall J.~N, 1980, ApJ, 235, 694
\bibitem{15} Becker, R.H., White, R.L., \& Helfand, D.J. 1995, ApJ, 450, 559
\bibitem{15}  Best P. N., R\"ottgering H. J. A., Longair, M. S., 2000, MNRAS, 311, 23
\bibitem{97}  Bolzonella, J.M. Miralles, R. Pell, 2000, A\&A, 363, 476
\bibitem{20}  Bruzual A.G., Charlot S., 2003, ApJ, 344, 1000
\bibitem{97}  Collister A.A., Lahav O., 2004, astro-ph/0311058
\bibitem{97}  Condon J.J., Cotton W.D., Greisen E.W., Yin Q.F., Perley R.A., Taylor G.B., Broderick J.J., 1998, AJ, 115, 1693
\bibitem{4}   Dunlop J.S., Peacock J.A., 1990, MNRAS, 247, 19 
\bibitem{4}   Eales S., Rawlings S., Law-Green D., Cotter G., Lacy M., 1997, MNRAS, 291, 593
\bibitem{249} Fanaroff B.L., Riley J.M., 1974, MNRAS, 167, 31
\bibitem{60}  Jackson C.A., Wall J.V., 1999, MNRAS, 304, 160
\bibitem{25} Inskip K.J., Best P.N., Rawlings S., Longair M.S., Cotter G., R\"ottgering H.J.A., Eales S., 2002, MNRAS, 337, 1381
\bibitem{25} Ivezi{\'c} Z. et al., 2002, AJ, 124, 2364
\bibitem{82}  Jarvis M.J., Rawlings S., 2000, MNRAS, 319, 121
\bibitem{82}  Jarvis M.J., Rawlings S., Eales S.A., Blundell K.M., Bunker A.J., Croft S., McLure R.J., Willott C.J., 2001, MNRAS, 326, 1585
\bibitem{9}   Laing R.A., Riley J.M., Longair M.S., 1983, MNRAS, 204,151
\bibitem{33}  Longair M.S., 1966, MNRAS, 133, 421
\bibitem{15} McLure R.~J., Willott C.J., Jarvis M.J., Rawlings S., Hill G.J., Mitchell E., Dunlop J.S., Wold M., 2004, MNRAS, submitted.
\bibitem{15}  Petrosian V., 1976, ApJ, 209, L1
\bibitem{67}  Rengelink, R.B., et al., 1997, AAPS, 124, 259
\bibitem{55}  Rowan-Robinson M.M., 1968, MNRAS, 138, 445
\bibitem{54}  Schmidt M., 1968, ApJ, 151, 393
\bibitem{9}   Snellen I.A.G., Best P.N., 2001, MNRAS, 328, 897
\bibitem{15}  Strauss M.A. et al., 2002, AJ, 124, 1810 
\bibitem{15}  Waddington I., Dunlop J.S., Peacock J.A., Windhorst R.A., 2001, MNRAS, 328, 882 
\bibitem{15}  White, R.L., Becker, R.H., Helfand, D.J., \& Gregg, M.D. 1997, ApJ, 475, 479
\bibitem{13}  Willott C.J., Rawlings S., Blundell K.M., Lacy M., 1999, MNRAS, 309, 1017
\bibitem{13}  Willott C.J., Rawlings S., Blundell K.M., Lacy M., Eales S.A., 2001, MNRAS, 322, 536
\bibitem{15}  Willott C.J., Rawlings S., Jarvis M.J., Blundell K.M., 2003, MNRAS, 339, 173
\end{thebibliography}
\end{document}